\documentclass[twocolumn,showpacs,preprintnumbers,amsmath,amssymb,prb, superscriptaddress]{revtex4}
\usepackage{graphicx}% Include figure files
\usepackage{dcolumn}% Align table columns on decimal point
\usepackage{bm}% bold math
\usepackage{amsmath, amssymb, graphics, setspace}
\usepackage{color}

\newcommand{\mathsym}[1]{{}}
\newcommand{\unicode}[1]{{}}

\begin{document}

%\preprint{APS/123-QED}

\title{Possible pairing symmetries in SrPtAs with a local lack of inversion center}% Force line breaks with \\

\author{Jun Goryo}\email{goryo@phys.ethz.ch}
\affiliation{Institute for Theoretical Physics, ETH Zurich, 8093 Zurich, Switzerland}
\affiliation{Institute of Industrial Science, the University of Tokyo, 153-8505 Tokyo, Japan}
\author{Mark H. Fischer}%
\affiliation{Department of Physics, Cornell University, Ithaca, NY 14850 USA}
\author{Manfred Sigrist}%
\affiliation{Institute for Theoretical Physics, ETH Zurich, 8093 Zurich, Switzerland}

\date{\today}% It is always \today, today,

\begin{abstract}
We discuss possible pairing symmetries in the hexagonal pnictide superconductor SrPtAs.
The local lack of inversion symmetry of the two distinct conducting layers in the unit cell results 
in a special spin-orbit coupling with a staggered structure. We classify the pairing symmetry by the global crystal point group $D_{3d}$, and 
suggest some candidates for the stable state using a tight-binding model with an in-plane, density-density type pairing interaction. 
We may have some unconventional states like $s+f$-wave and a mixture of 
chiral $d$-wave and chiral $p$-wave. The spin orbit coupling is larger than the interlayer hopping, and 
the mixing between spin-singlet and triplet states can be seen in spite of the fact that the system has a global inversion center. 

%We discuss possible pairing symmetries in the hexagonal pnictide superconductor SrPtAs.
%The local lack of inversion symmetry in the two distinct conducting layers of the unit cell results in a special spin-orbit coupling with a staggered structure. 
%We analyze the stable pairing symmetry within a tight-binding model with an in-plane, density-density type pairing interaction and inter-band pair scatterings. 
%Due to the locally anti-symmetric spin orbit coupling and weak interlayer hoppings suggested by the band structure calculation, the mixing between 
%spin-singlet and triplet states can be seen in spite of the fact that the system has global inversion symmetry. 
%When the on-site attractive interaction is reduced or becomes repulsive, we may have some unconventional pairing states like $s+f$-wave or a mixture of 
%chiral $d$-wave and chiral $p$-wave, which appear in the group theoretical classification under the global crystal symmetry $D_{3d}$. 
%For sufficiently strong nearest-neighbor interactions,
%we find that a mixture of states with chiral $d$- and chiral $p$-wave pairing symmetries overcomes the conventional $s$-wave pairing state. 
%Namely, this state exhibits spin-singlet and spin-triplet mixing 
%due to the local lack of inversion symmetry, and according to a forth-order Ginzburg-Landau analysis, breaks time-reversal symmetry. 
%Finally, we also discuss topological features of this unconventional state, which supports non-trivial surface states.   
\end{abstract}

\maketitle

%\section{Introduction}
%Since the proposal and subsequent discovery of the two-dimensional topological insulator, topological states of matter have attracted much attention in recent years. In addition to topological insulator - a system with a non-trivial insulating bulk - also the possibility of having topological superconductors has been discussed. As in the topological insulators, bulk excitations are gapped in a non-trivial way. A common feature for these new states of matter is a strong spin-orbit coupling. A possible realization for such a state is the Cu doped topological insulator Bi$_{2}$Se$_{3}$, Cu$_{x}$Bi$_{2}$Se$_{3}$. 

%In this work, we propose another candidate material to possess a topological superconducting state, 

The relation between crystal structure and pairing symmetry plays an important role in unconventional superconductivity.\cite{Sigrist-Ueda} 
Pairing states can be categorized with respect to the irreducible representations of the point group of the crystal lattice and do not mix unless they belong to the same representation. 
%The invariant representation allows spin-singlet pairing only, whereas both of the spin-singlet and spin-triplet states are possible in non-trivial representations.   
Since the Pauli principle requires that the momentum part of 
singlet and triplet states possess even and odd parity, respectively, 
their mixing is prohibited in a system with inversion symmetry. Superconductivity in non-centrosymmetric 
systems, i.e., CePt$_3$Si, opens however the possibility of singlet-triplet mixing.\cite{Gorkov-Rashba, Bauer,Frigeri}
It plays a key role to explain the puzzling behavior of the observed nuclear spin-lattice relaxation rate $T_1^{-1}$.\cite{Hayashi_etal} 
Microscopically, this mixing is caused by an anti-symmetric spin-orbit coupling (SOC). 

Recently, possible singlet-triplet mixing in centrosymmetric systems with a local lack of inversion symmetry, such as special crystal lattices or heterostructures, was discussed.\cite{Fischer, Maruyama-Yanase} 
%We may ask if the singlet-triplet mixing can be seen in a system with a local lack of inversion symmetry.~\cite{Maruyama-Yanase}
The recently-discovered hexagonal pnictide superconductor SrPtAs\cite{SrPtAsexp} ($T_c=2.4K$) belongs to the former case of a special crystal structure. The unit cell possess a global inversion center and its point group is $D_{3d}$. There are two distinct honeycomb Pt-As layers within the unit cell each of which has  no inversion center. 
LDA calculations revealed that these two layers are conducting 
with only a small inter-layer hopping, i.e., the system is quasi-two-dimensional (quasi-2D). %\cite{SrPtAs,SrPtAs2} 
In addition, a large splitting of the bands due to anti-symmetric spin-orbit coupling (SOC) was seen. The consequences of this local lack of inversion symmetry on magnetic properties of the superconducting phase\cite{SrPtAs} as well as on
electronic phenomena\cite{SrPtAs2} has previously been studied. In this work, we aim at clarifying its role for the pairing symmetry. 

Table \ref{classification} shows the classification of the pairing states based on the global symmetry of the crystal $D_{3d}$. 
We assume intra-layer pairing due to the quasi-2D nature of the system, and focus on on-site and nearest-neighbor-site (nn-site) pairing interactions. 
%As sites are neglected, since Pt 5d orbital is dominant in the metallic state.\cite{SrPtAs,SrPtAs2} The result is shown in Table \ref{classification}. 
It is intriguing that in this table both even-parity spin-triplet and odd-parity spin-singlet pairing appear.  
%(Note that parity is conserving, since the crystal has an inversion center). 
The reason is that we have two distinct layers in the unit cell 
indicated by $l=1,2$, and we can introduce an odd-parity factor $(-1)^l$ under the global inversion operation. 
Multiplying this factor to a certain pair wave function results in even-parity spin-triplet or odd-parity spin-singlet states. 
Moreover, spin-singlet and triplet states coexist in some irreducible representations, namely $A_{1g}$, $E_g$, 
$A_{2u}$ and $E_u$. Therefore, mixing of spin-singlet and triplet states becomes possible in these representations despite the parity conservation.

%%%%%%%%%%%%%%%%
%%%%%%%%%%%%%%%%
\begin{table*}
\caption{(a) Spin-singlet, and (b) spin-triplet basis gap functions. This classification is based on $D_{3d}$ symmetry. The index $l=1,2$ denotes two distinct layers. 
The definitions for functions of crystal momentum 
$\bm k$ are, $e_{\bm k}\equiv \sum_n \cos \bm k \cdot \bm T_n$, $e^+_{\bm k}\equiv \sum_n \omega^n \cos \bm k \cdot \bm T_n$,  $o_{\bm k}\equiv \sum_n \sin \bm k \cdot \bm T_n$, $o^+_{\bm k}\equiv \sum_n \omega^n \sin \bm k \cdot \bm T_n$, $e^-_{\bm k}=e^{+*}_{\bm k}$,$o^-_{\bm k}=o^{+*}_{\bm k}$, where ${\bm T}_{n=1,2,3}$ is the bond vector between nearest-neighbor sites,  
%with $\bm T_1=(0,a,0)$, $\bm T_2=(\sqrt{3}a/2,-a/2,0)$, and $\bm T_3=(-\sqrt{3}a/2,-a/2,0)$ the bond vectors between nearest-neighbor Pt sites in a plane 
%($a$ is the in-plane lattice constant), 
and $\omega^{n}=\exp[2 n \pi i /3]$.  
Note that we have even-parity spin-triplet part and odd-parity spin-singlet part due to the odd-parity factor $(-1)^l$.}
\begin{ruledtabular}
\begin{tabular}{llll}
$\Gamma$ & Parity & (a) spin-singlet & (b) spin-triplet 
\\
&& $\hat{\Delta}_{\bm k l}^{\Gamma,m}=i\hat\sigma_y\psi_{\bm k l}^{\Gamma,m}$ &  $\hat{\Delta}_{\bm k l}^{\Gamma,m}=i[\hat{\bm \sigma} \cdot {\bm d}_{\bm k l}^{\Gamma,m}]\hat\sigma_y$
\\\hline
$A_{1g}$ 
&
& $\psi_{l}^{{A}_{1g}}=1$, $\psi_{\bm k l}^{{A}_{1g}}=e_{\bm k}$
& ${\bm d}_{\bm k l}^{{A}_{1g}}=(-1)^lo_{\bm k}\hat{\bm z}$
\\
$A_{2g}$ 
& Even
&
& ${\bm d}_{\bm k l}^{{A}_{2g}}=(-1)^lo_{\bm k}\hat{\bm x}_{\pm}$
\\
$E_g$ 
&
& $\psi_{\bm k l}^{{E}_g,1}=e^+_{\bm k}$
& ${\bm d}_{\bm k l}^{{E}_g,1}=(-1)^lo^+_{\bm k}\hat{\bm z}$
\\
&
& $\psi_{\bm k l}^{{E}_g,2}=e^-_{\bm k}$
& ${\bm d}_{\bm k l}^{{E}_g,2}=(-1)^lo^-_{\bm k}\hat{\bm z}$
\\
\hline
$A_{1u}$ 
&
& 
&${\bm d}_{\bm k l}^{{A}_{1u}}=o_{\bm k}\hat{\bm x}_{\pm}$
\\
$A_{2u}$ 
& Odd 
& $\psi_l^{{A}_{2u}}=(-1)^l$, $\psi_{\bm k l}^{{A}_{2u}}=(-1)^le_{\bm k}$
&${\bm d}_{\bm k l}^{{A}_{2u}}=o_{\bm k}\hat{\bm z}$
\\
$E_u$ 
&
& $\psi_{\bm k l}^{{E}_u,1}=(-1)^le^+_{\bm k}$
& ${\bm d}_{\bm k l}^{{E}_u,1}=o^+_{\bm k}\hat{\bm z}$
\\
&
& $\psi_{\bm k l}^{{E}_u,2}=(-1)^le^-_{\bm k}$
& ${\bm d}_{\bm k l}^{{E}_u,2}=o^-_{\bm k}\hat{\bm z}$
\end{tabular}
\end{ruledtabular}
\label{classification}
\end{table*}
%%%%%%%%%%%%%%%%
%%%%%%%%%%%%%%%%

\begin{figure}
\begin{center}
\includegraphics[width=7cm]{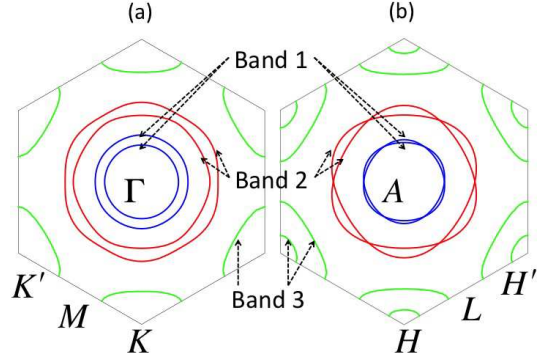}
\end{center}
\caption{Fermi surfaces at (a) $k_z=0$, and (b) $k_z=\pi/c$. Inner blue, middle red, and outer green lines show the Fermi surfaces for 
band 1, band 2, and band 3, respectively. Note that there are two branches in each band as suggested in Eq. (\ref{normal_spectrum}), but one of the branches 
in band 3 does not cross the Fermi level at $k_z=0$.}
\label{FS}
\end{figure}

%We are able to discuss the phenomenological identification of pairing symmetry by using this classification table.\cite{Sigrist-Ueda} 
%Unfortunately, at present, 
Since there is no experimental information on the pairing symmetry at present, 
%due to the fact that a single crystal has not been produced yet. We thus build a proto-type model for the pairing and 
we discuss some potential candidates for the stable symmetry within a simple model. 
We use a tight-binding description for electrons on the Pt sites with a Hamiltonian consisting of two parts: 
$H=H_0+H_{sc}$. 
%-g_{on}\sum_{i,l,b} n_{ilb\uparrow} n_{ilb\downarrow}
%-g_{nn}\sum_{\langle i,j \rangle,b, l}\sum_{s,s^\prime} n_{ilbs} n_{jlbs^\prime}
The first part, $H_0$, is the one-body Hamiltonian introduced by Refs.~\onlinecite{SrPtAs} and \onlinecite{SrPtAs2} in order to reproduce the LDA band structure of SrPtAs,
\begin{equation}
H_0=\sum_{\bm k,l,l^\prime, s,b}\epsilon^{(b)}_{\bm k ll^\prime}c^{(b)\dagger}_{\bm k ls}c^{(b)}_{\bm k l^\prime s}
+\sum_{\bm k, l, s,b} \alpha_{b} {\bm \lambda}_{\bm k l}\cdot \bm \sigma_{ss^\prime}c^{(b)\dagger}_{\bm k l s}c^{(b)}_{\bm k l s^\prime},
\label{eq:ham0}
\end{equation}
with 
\begin{eqnarray}
\epsilon^{(b)}_{\bm k ll^\prime}&=&(\epsilon^{(b)}_{1\bm k}-\mu_b){\tau}^{0}_{ll^\prime}+{\rm Re}[\epsilon^{(b)}_{c\bm k}]{\tau}^{1}_{ll^\prime}+{\rm Im}[\epsilon^{(b)*}_{c\bm k}]{\tau}^{2}_{ll^\prime},
\nonumber\\
%{\bm \lambda}_{\bm k l}&=&(-1)^l\hat{\bm z}\sum_{i=1,2,3} \sin \bm k \cdot {\bm T}_i \equiv (-1)^l {\bm \lambda}_{\bm k},
{\bm \lambda}_{\bm k l}&=&(-1)^l {\bm \lambda}_{\bm k} = {\tau}^3_{ll} {\bm \lambda}_{\bm k},
\label{eq:ham}
\end{eqnarray}
where $c^{(b)}_{\bm k l s}$ ($c^{(b)\dagger}_{\bm k l s}$) is the annihilation (creation) operator of an electron in the $b$-th band ($b=1,2,3$) with 
crystal momentum $\bm k$, 
spin $s$ in the $l$-th layer ($l=1,2$). In the above equation, we introduced $\hat\sigma^0$ ($\hat\tau^0$) and $\hat\sigma^i$ ($\hat\tau^i$), the unit and Pauli matrices acting on the spin (layer) space. Including Pt nearest-neighbor hopping within the plane, as well as nearest- and next-nearest-neighbor hopping between the planes, one finds 
$\epsilon^{(b)}_{1 \bm k}=t^{(b)}_{1} \sum_n \cos {\bm k}\cdot {\bm T}_n+t^{(b)}_{c2}\cos (c k_z)$, and $\epsilon^{(b)}_{c \bm k}=t^{(b)}_{c} \cos (k_z c/2)[1+\exp(-i\bm k \cdot \bm T_3)+\exp(i\bm k \cdot \bm T_2)]$ with 
$\bm T_1=(0,a,0)$, $\bm T_2=(\sqrt{3}a/2,-a/2,0)$, and $\bm T_3=(-\sqrt{3}a/2,-a/2,0)$ the 
in-plane nearest-neighbor bond vectors used in the tight-binding approach ($a$ and $c$ are in-plane and inter-layer lattice constants).
An important ingredient is the locally anti-symmetric SOC ${\bm \lambda}_{\bm k l}$, which reads 
${\bm \lambda}_{\bm k}=\hat{\bm z}\sum_n \sin \bm k \cdot {\bm T}_n$ for each band.
This term is symmetric under global inversion, but anti-symmetric under the local inversion operation in each layer. 
Due to the Kramers degeneracy, there are only two branches in the energy spectrum of the Hamiltonian \eqref{eq:ham0} for each band
\begin{eqnarray}
\xi^{(b)}_{\bm k \pm}=\epsilon^{(b)}_{1\bm k}-\mu_b \pm \sqrt{|\epsilon^{(b)}_{c \bm k}|^2+|\alpha_b {\bm \lambda}_{\bm k}|^2}.
\label{normal_spectrum}
\end{eqnarray}
%Table~\ref{tab:tight-binding} lists the tight-binding parameters used in the following for numerical calculations. 
We use tight-binding parameters from Ref. \onlinecite{SrPtAs} which lead to Fermi surfaces as shown in Fig.~\ref{FS}.  
With this parameters, the outermost band, labelled band 3, is the dominant band with 74$\%$ of the total density of states (DOS) due to its proximity to the van Hove singularity (vHS) at the $M$ points in the Brillouin zone (BZ). 
%This large fraction arises due to proximity of the band to the saddle points 
%along M-L line in the Brillouin zone (BZ), see Fig.~\ref{FS}. 
Note that the ratio $\alpha_b/t^{(b)}_{c}$, which parametrizes the effect of 
the local lack of inversion symmetry, is comparable or larger than 1. This large ratio plays an essential role for the mixing between spin-singlet and spin-triplet state, as we will see below.   
%%%%%%%%%%%%%%%%
%%%%%%%%%%%%%%%%
%\begin{table}
%\caption{Tight-binding parameters and the fraction of the density of states (DOS) for three bands which provide Fermi surfaces. The parameters 
%are given in the unit of the nearest-neighbor hopping integral in band 2. Due to the SOC $\alpha_b$, each band has split Fermi surfaces depicted in Fig. \ref{FS}.}
%\begin{ruledtabular}
%\begin{tabular}{cccccc|c}
%$b$ & $t^{(b)}_{1}$ & $t^{(b)}_{c}$ & $t^{(b)}_{c2}$ & $\alpha_b/t^{(b)}_c$ & $\mu_b$ & DOS 
%\\\hline
%band 1 & 1.25&0.1 & 0.05& 4 & 2.5 & 9$\%$
%\\
%band 2 & 1& 0.1& 0.05& 2.8& 0.5 & 17$\%$
%\\
%band 3 & -0.48& 0.075& -0.03& 0.6& 0.6 & 74$\%$
%\end{tabular}
%\end{ruledtabular}
%\label{tab:tight-binding}
%\end{table}
%%%%%%%%%%%%%%%%
%%%%%%%%%%%%%%%%
%%%%%%%%%%%%%%%%
%%%%%%%%%%%%%%%%
%%%%%%%%%%%%%%%%
%%%%%%%%%%%%%%%%

For the pairing term $H_{sc}$ in the total Hamiltonian we assume intra-layer interactions 
including density-density type attractive interaction, as well as inter-band pair scatterings allowed by the kinematics.   
%\begin{widetext}
%\begin{multline}
%H_{sc}=
%-\sum_{\bm k,\bm k^\prime}\sum_{l,b,b^\prime}V_{on}^{bb^\prime}c^{(b)\dagger}_{\bm k l \uparrow } c^{(b)\dagger}_{-\bm k l \downarrow} c^{(b^\prime)}_{-\bm k^\prime l \downarrow} 
%c^{(b^\prime)}_{\bm k^\prime l \uparrow}
%\label{interaction}
%\\
%-\sum_{\bm k,\bm k^\prime,s,s^\prime}\sum_{l,b,b^\prime} 
%\frac{1}{2}V_{nn;\bm k,\bm k^\prime}^{bb^\prime} c^{(b)\dagger}_{\bm k l s} c^{(b)\dagger}_{-\bm k l s^\prime} c^{(b^\prime)}_{-\bm k^\prime l s^\prime} c^{(b^\prime)}_{\bm k^\prime l s },
%\end{multline}
%\end{widetext}
%where $V_{nn;\bm k,\bm k^\prime}^{bb^\prime}=V_{nn}^{bb^\prime}\sum_i \cos[(\bm k-\bm k^\prime)\cdot \bm T_i] $, and the first and the second terms are on-site and nearest-neighbor pairing 
%interactions including the inter-band scattering, respectively. 
Using the basis functions from Table \ref{classification}, $H_{sc}$ is written in Fourier form as 
\begin{eqnarray}
H_{sc}=\sum V^{bl;l^\prime b^\prime}_{s_1 s_2; s_3 s_4}(\bm k, \bm k^\prime)c^{(b)\dagger}_{-\bm k l s_1} c^{(b)\dagger}_{-\bm k l s_2} 
c^{(b^\prime)}_{-\bm k^\prime l^\prime s_3} c^{(b^\prime)}_{\bm k^\prime l^\prime s_4}
\label{interaction}
\end{eqnarray}
with
\begin{multline}
V^{bl; l^\prime b^\prime }_{s_1 s_2; s_3 s_4}(\bm k, \bm k^\prime)=-g^{bb^\prime}_{\rm on}\sum_{\Gamma} \psi_l^{(\Gamma)}\psi_{l^\prime}^{(\Gamma)*} (\hat\sigma_y)_{s_1s_2} (\hat\sigma_y)_{s_3s_4}
\\
-g^{bb^\prime}_{\rm nn}\sum_{\Gamma,m}\psi_{\bm k l}^{(\Gamma,m)}\psi_{\bm k^\prime l^\prime}^{(\Gamma,m)*}  (\hat\sigma_y)_{s_1s_2} (\hat\sigma_y)_{s_3s_4}
\\
-g^{bb^\prime}_{\rm nn}\sum_{\Gamma,m}d_{i \bm k l}^{(\Gamma,m)}d_{j \bm k^\prime l^\prime}^{(\Gamma,m)*} (\hat\sigma_i \hat\sigma_y)_{s_1s_2} (\hat\sigma_y \hat\sigma_j)_{s_3s_4}, 
\end{multline}
where $g^{bb^\prime}_{\rm on}$ and $g^{bb^\prime}_{\rm nn}$ are the 
coupling constants for on-site and nearest-neighbor channels.  
%We expext the pairing instability to occur in band 3 with its dominant contribution to the DOS. 
The pairing instability in this model occurs in band 3 with its dominant contribution to the DOS. 
Smaller gaps then open on the other two bands due to pair scattering.
%Our model discussion shows that the pairing naturally occurs in the dominant band 3, and that the inter-band 
%pair scattering opens also gaps on the other bands. 
%We do not take into account  any inter-layer pairing interactions as we mentioned before. 
 
We solve the linearized gap equation (the eigenvalue equation for $T_c$) 
\begin{multline}
\Delta^{(b)}_{s_1s_2}(\bm k)=
-T_c \sum_{\bm k', \omega_n} V^{bl; l^\prime b^\prime}_{s_1 s_2; s_3 s_4}(\bm k, \bm k^\prime)\\
\times[\hat{G}^{(b^\prime)}(\bm k^\prime, i\omega_n) \hat\Delta^{(b^\prime)}(\bm k^\prime)
\hat{G}^{(b^\prime)}(-\bm k^\prime, -i\omega_n)]_{l^\prime k^\prime}^{s_3s_4}, 
\label{linearized-gap-eq}
\end{multline}
where the sum runs over repeated indices, and 
\begin{eqnarray}
\hat{G}^{(b)} (\bm k, i \omega_n)=
\left\{\hat\sigma_0\otimes(i \omega_n \hat\tau^0 + \hat\epsilon^{(b)}_{\bm k})+\alpha_b {\bm \lambda}_{\bm k} \cdot \hat{\bm \sigma}\otimes\hat\tau^3\right\}^{-1}
\end{eqnarray}
is the normal-state Matsubara Green's function. 
All the possible gap functions are listed as 
\begin{eqnarray}
&&\hat\Delta_{\bm k l}^{(b)\Gamma}=
\label{gap_function}\\
&&\left\{
\begin{array}{ll}
\hat\Delta_\Gamma^{(b)}( \psi_l^{\Gamma}+s^{(b)}_{\Gamma}\psi_{\bm k l}^{\Gamma}+t^{(b)}_\Gamma \bm d_{\bm k l}^{\Gamma}\cdot \hat{\bm \sigma} )i \hat\sigma_y & \Gamma=A_{1g}
\\
\\
\Delta_\Gamma^{(b)} \bm d_{\bm k l}^{\Gamma}\cdot i \hat{\bm \sigma} \hat\sigma_y & \Gamma=A_{2g}
\\
\\
\sum_m \Delta^{(b)}_{\Gamma,m} \left(\psi_{\bm k l}^{\Gamma,m}+t^{(b)}_{\Gamma} \bm d_{\bm k l}^{\Gamma,m}\cdot \hat{\bm \sigma}\right) i \hat\sigma_y & \Gamma=E_g
\\
\\
\Delta_\Gamma^{(b)} \bm d_{\bm k l}^{\Gamma}\cdot i \hat{\bm \sigma} \hat\sigma_y & \Gamma=A_{1u}
\\
\\
\Delta_\Gamma^{(b)}(\tilde{s}^{(b)}_{\Gamma}\psi_l^{\Gamma}+s^{(b)}_{\Gamma}\psi_{\bm k l}^{\Gamma}+{\bm d}_{\bm k l}^{\Gamma}\cdot \hat{\bm \sigma} )i \hat\sigma_y & \Gamma=A_{2u}
\\
\\
\sum_m \Delta^{(b)}_{\Gamma,m} \left(s^{(b)}_{\Gamma} \psi_{\bm k l}^{\Gamma,m}+\bm d_{\bm k l}^{\Gamma,m}\cdot \hat{\bm \sigma}\right) i \hat\sigma_y & \Gamma=E_u
\end{array}
\right.
\nonumber
\end{eqnarray}
where $\Delta^{(b)}_\Gamma$ and $\Delta^{(b)}_{\Gamma,m=1,2}$ are the order parameters, and $s^{(b)}_{\Gamma}$ and $t^{(b)}_{\Gamma}$ are the mixing ratios 
of subdominant spin-singlet and triplet parts, respectively. 
We see in $\Gamma=A_{1g} $ and $A_{2u}$ that there is a mixing between on-site and nearest-neighbor-site pairings, besides the spin-singlet and triplet mixing. 
%To solve eq. (\ref{linearized-gap-eq}), we obtain $T_c$ and mixing ratios for each $\Gamma$. 
%The state with ${\max}(T_c)\equiv T_c^{\max}$ is the most stable state. 
We neglect the band dependence of the intra-band couplings, namely, 
$g_{\rm on(nn)}=g_{\rm on(nn)}^{1,1}=g_{\rm on(nn)}^{2,2}=g_{\rm on(nn)}^{3,3}$, and
introduce repulsive inter-band interactions $g_{\rm on(nn)}^{1,3}=g_{\rm on(nn)}^{2,3}=-0.05$, 
keeping $g_{\rm on(nn)}^{1,2}=0$. This choice is motivated by the nesting-like 
structures between band 2 and 3, and band 1 and 3, respectively.\cite{FeAs} 
 We can then calculate 
the state with the maximum eigenvalue $T_c^{\max}$ at a point $(g_{\rm on}, g_{\rm nn})$ in 
the coupling constant space. 

\begin{figure}
\begin{center}
\includegraphics[width=7.5cm]{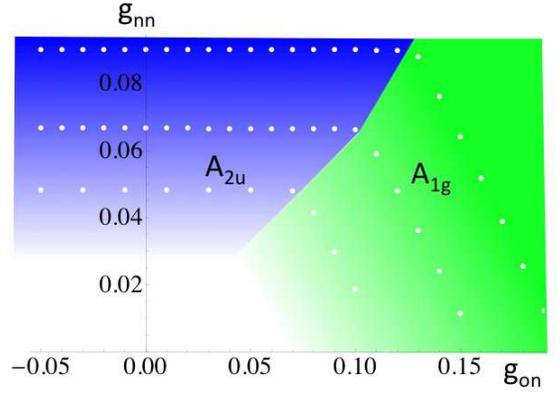}%
\end{center}
\caption{The phase diagram of the stable pairing states in the coupling constant space $(g_{\rm on},g_{\rm nn})$. The tight-binding parameters 
suggested by LDA calculation\cite{SrPtAs,SrPtAs2} is used. The sequences of dots show equal $T_c$ lines 
at $T_c/t_1^{(2)}=10^{-5}, 10^{-4}, 10^{-3}$ from bottom to top.
%The stable states and their $T_c=T_c^{\max}$ at $(g_{\rm on},g_{\rm nn})$. $A_{1g}$ state is stable at the 
%points indicated by green circles, and $A_{2u}$ state by blue squares. The lowest, middle, and top sequences of dots show equal $T_c$ lines
%at 1K, 5K, and 10K, estimated with assuming the nearest-neighbor hopping in band2 $t_1^{(2)}=1 eV$. 
}
\label{phase-diagram1}
\end{figure}

%We obtain the phase diagram shown in 
Figure \ref{phase-diagram1} shows the obtained phase diagram.  
The $A_{1g}$ state is stabilized when the on-site attraction is dominant, whereas the $A_{2u}$ state 
becomes stable in the parameter region where the nn-site attraction is comparable to, or larger 
than the on-site coupling.  From Table \ref{classification} and Eq. (\ref{gap_function}), we see that both, the 
$A_{1g}$ and the $A_{2u}$ state, have ``$s+f$''-wave pairing symmetry, with the  
$s$-wave ($f$-wave) component dominant while the $f$-wave ($s$-wave) component with an odd-parity factor $(-1)^l$ is subdominant. 
Therefore, the quasiparticle excitations are fully gapped in the $A_{1g}$ state, whereas  line nodes appear in the $A_{2u}$ state. The $A_{2u}$ state invokes a full
coherence factor due to the $s$-wave component, and would show both Hebel-Slichter peak and power-law type temperature dependence of 
$T_1^{-1}$ like CePt$_3$Si.\cite{Hayashi_etal}
The gap structure involves sign changes which give rise to zero-energy Andreev bound states at certain surfaces, e.g. for the normal vector [010].\cite{YTanaka}  Note that 
the relation of the bound state and topology of the wave function has been discussed in Refs.~\onlinecite{MSato,Schnyder}. This state belongs to the class AIII of the topological classification\cite{Schnyder-Ryu-Furusaki-Ludwig}. 

%%%%%%%%%%%%%%%%
%%%%%%%%%%%%%%%%
%%%%%%%%%%%%%%%%
%%%%%%%%%%%%%%%

The locally anti-symmetric SOC introduces a mixing between spin-singlet and triplet parts, which is 
proportional to 
\begin{eqnarray}
\sum_{\bm k l} \frac{\psi^{\Gamma*}_{\bm k l}\left\{\bm d^{\Gamma}_{\bm k l} \cdot \alpha_b{\bm \lambda}_{\bm k}\right\}}{\sqrt{|\epsilon^{(b)}_{c\bm k}|^2+\alpha_b^2\lambda_{\bm k}^2}}
\left(\frac{1}{\xi^{(b)}_{\bm k+}}\tanh\frac{\xi^{(b)}_{\bm k+}}{2 T_c}-\frac{1}{\xi^{(b)}_{\bm k-}}\tanh\frac{\xi^{(b)}_{\bm k -}}{2 T_c}\right). 
\label{mixing}
\end{eqnarray}
This suggests that the mixing is suppressed by a large inter-layer hopping, as expected,  
since the system has global inversion symmetry and 
the locally anti-symmetric nature is smeared out when the three dimensionality becomes strong. Such a behavior is also seen in the magnetic susceptibility.\cite{SrPtAs} In this system, however, the inter-layer hopping has been estimated to be comparable or smaller than the SOC\cite{SrPtAs,SrPtAs2} and we hence expect a finite value of 
mixing. Indeed, around the boundary between the $A_{1g}$ and $A_{2u}$ phases in Fig.~\ref{phase-diagram1}, we find enhanced mixing ratios. 
Their magnitudes are almost band-independent and typical values are  $(s^{(b)}_{A_{1g}},t^{(b)}_{A_{1g}})=(-0.51,0.12)$ in the $A_{1g}$ phase, and 
$(\tilde{s}^{(b)}_{A_{2u}},s^{(b)}_{A_{2u}})=(0.15,-0.18)$ in the $A_{2u}$ phase (definitions of the ratios are given in Eq.~(\ref{gap_function})). 

\begin{figure}
\begin{center}
\includegraphics[width=7.5cm]{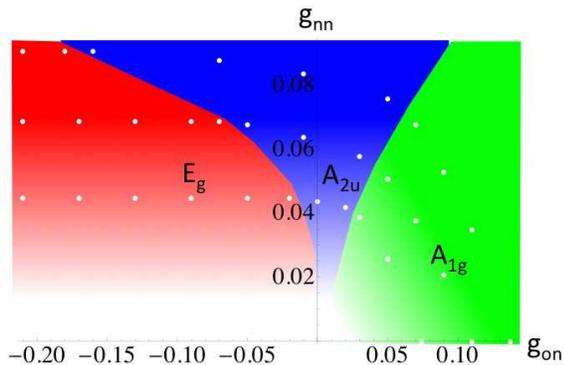}
\end{center}
\caption{The phase diagram of the stable paring state in the coupling constant space $(g_{\rm on},g_{\rm nn})$ in 
the DOS enhanced situation, where the Fermi level is located at the vHS point of band 3. The sequences of dots show equal $T_c$ lines 
at $T_c/t_1^{(2)}=10^{-5}, 10^{-4}, 10^{-3}$ from bottom to top. }
\label{phase-diagram2}
\end{figure}

Figure \ref{phase-diagram2} shows the phase diagram for a shifted chemical potential such that band 3 approaches the vHS. The enhanced DOS naturally leads to reduced coupling constants for the same $T_c$ as compared to the previous situation. More remarkably, 
%We now consider the situation of an enhance DOS in band 3 by shifting the Fermi energy in a way as to approach the van Hove singularity (vHS). The vHS is located between the corners of BZ boundary. 
%The result of such a shift is shown in Fig. \ref{phase-diagram2}. We examined whether  
%the coupling constants at the same $T_c$ are reduced compared to the previous result due to the enhancement of the DOS. 
%The remarkable thing is that 
the $E_g$ state shows up in the region where the on-site coupling is repulsive.  
One of the reasons for its stability is that the amplitude of the singlet component $|\psi^{E_g,m}_{\bm k,l}|$ 
has peaks at the saddle points, which is compatible with the Fermi surface structure. This phase involves two degenerate basis states 
indicated by $m=1,2$ in Eq. (\ref{gap_function}), 
and they make up a Kramers pair. A fourth-order analysis of the Ginzburg-Landau theory yields to degenelate states 
%tells us that the state has two-fold degeneracy 
$(\Delta^{(b)}_{E_g,1},\Delta^{(b)}_{E_g,2})=(1,0),(0,1)$, which both break time-reversal symmetry.
%is broken spontaneously. 
We focus here on the 
first configuration and set $\Delta^{(b)}_{E_g,2}=0$. Expanding the spin-singlet component around the zone-central 
axes $k_x=k_y=0$ gives $\psi^{E_g,1}_{\bm k,l}=(k_x+ik_y)^2$ with $d_{x^2-y^2}+id_{xy}$-wave 
symmetry, or chiral $d$-wave symmetry. Note that $d_{x^2-y^2}$ and $d_{xy}$ components are degenerated 
in the three-fold rotational symmetry. The same expansion for the spin-triplet part yields ${\bm d}^{E_g,1}_{\bm k l}=(-1)^l(k_x-ik_y)\hat{\bm z}$ with 
chiral $p$-wave symmetry like Sr$_2$RuO$_4$.\cite{Mackenzie-Maeno} The chiral $d$-wave part has $L_z=+2$, whereas the chiral $p$-wave part 
$L_z=-1$ ($L_z$: $z$-component of the relative angular momentum of the pair). 
These states can mix with each other as indicated by Table \ref{classification}. \footnote{Indeed, the eigenvalues for the three-fold rotation $e^{2 \pi i L_z /3 }$ are the same.} 
The mixed state is classified into class A in the scheme of the topological classification.\cite{Schnyder-Ryu-Furusaki-Ludwig}
Due to the chiral nature of the pairing, this state has a non-zero value for the Chern number defined by the vorticity of the quasiparticle wave function in $k$ space\cite{TKNN,Kohmoto85} and supports chiral edge states topologically.\cite{Volovik,Read-Green} 

%We have checked that the stability of the state, especially the chiral $d$-wave component, is enhanced in the zero inter-layer hopping limit. It is reasonable, since there is no %$k_z$ dependence in the energy spectrum and Fermi surfaces do not move away from the vicinity of saddle points, where the chiral $d$-wave amplitude has peaks.  

%%%%%%%%%%%%%%%%
%%%%%%%%%%%%%%%%
%%%%%%%%%%%%%%%%
%%%%%%%%%%%%%%%

Our analysis provides insight into the basic trends of the hexagonal superconductor SrPtAs 
whose electrons experience a locally non-centrosymmetric environment. The $A_{1g}$ state is stable in the electron-phonon coupling limit  
where on-site attraction is dominant. On the other hand, in the strongly-correlated limit with on-site repulsion or strong nearest-neighbor attraction, the
$A_{2u}$ state is stabilized. In this state, line nodes coming from the spin-triplet component cause a power-law behavior of $T_1^{-1}$, whereas 
a Hebel-Slichter peak arises slightly below $T_c$ due to the coherence factor of the spin-singlet component, 
in analogy with CePt$_3$Si.\cite{Hayashi_etal} Such a behavior would be a strong signal of the 
locally anti-symmetric SOC. As mentioned, the nodal structure results in Andreev bound states at a certain surface,\cite{YTanaka} which is related to 
the topology of the bulk state.\cite{MSato,Schnyder} The $E_g$ state 
is possible in some particular cases like DOS enhanced situation owing to the vHS of the saddle points in the hexagonal BZ. 
This state breaks time-reversal symmetry whose signal could be detected by $\mu$SR measurement for spontaneous magnetization around impurities and also 
the Kerr rotation experiment, for examples. The state has chirality which is characterized by the Chern number, and leads to topologically-protected chiral edge states.\cite{Volovik,Read-Green} 

The authors are grateful to D.F. Agterberg, P. Brydon, A. Schnyder and G.-Q. Zheng for stimulating discussions. 
J.G. is financially supported by a Grant-in-Aid for Scientific Research 
from Japan Society for the Promotion of Science, Grant 
No. 23540437 and  by Yamada Science Foundation as well as the Pauli Center for Theoretical Studies of ETH Zurich.
MHF acknowledges support from NSF Grant DMR-0955822, as well as from NSF Grant DMR-0520404 to the Cornell Center for Materials Research.

%%%%%%%%%%%%%%%%
%%%%%%%%%%%%%%%%


\begin{references}

\bibitem{Sigrist-Ueda} M. Sigrist and K. Ueda, Rev. Mod. Phys. {\bf 63}, 239 (1991).

\bibitem{Gorkov-Rashba} L. P. Gor'kov and E. I. Rashba, Phys. Rev. Lett. {\bf 87}, 037004 (2001). 

\bibitem{Bauer} E. Bauer {\it et. al.}, Phys. Rev. Lett. {\bf 92}, 027003 (2004).

\bibitem{Frigeri} P. A. Frigeri, D. F. Agterberg, A. Koga, and M. Sigrist, Phys. Rev. Lett. {\bf 92}, 097001 (2004). 

\bibitem{Hayashi_etal} N. Hayashi, K. Wakabayashi, P. A. Frigeri, and M. Sigrist, Phys. Rev. B {\bf 73}, 092508 (2006). 

\bibitem{Fischer} M.H. Fischer, F. Loder, and M. Sigrist, Phys. Rev. B {\bf 84} 184533 (2011).

\bibitem{Maruyama-Yanase} D. Maruyama, M. Sigrist, and Y. Yanase, J. Phys. Soc. Jpn. {\bf 81} 034702 (2012). 

\bibitem{SrPtAsexp} Y. Yoshikubo, K. Kudo, and M. Nohara, J. Phys. Soc. Jpn. {\bf 80}, 055002 (2011). 

\bibitem{SrPtAs} S. J. Youn, M. H. Fischer, S. H. Rhim, M. Sigrist, and D. F. Agterberg, Phys. Rev. B {\bf 85}, 220505 (2012). 

\bibitem{SrPtAs2} S. J. Youn, S. H. Rhim, D. F. Agterberg, M. Weinert, and A. J. Freeman, arXiv:1202.1604

\bibitem{FeAs} Y. Kamihara, T. Watanabe, M. Hirano and H. Hosono, J. Am. Chem. Soc. {\bf 130}, 3296 (2008); I.I. Mazin and J. Schmalian, Physica C {\bf 469}, 614 (2009).

\bibitem{YTanaka} S. Kashiwaya and Y. Tanaka, Rep. Prog. Phys. {\bf 63}, 1641 (2000). 

\bibitem{MSato} M. Sato, Y. Tanaka,  K. Yada, and T. Yokoyama, Phys. Rev. B {\bf 83},  224511 (2011). 

\bibitem{Schnyder} A. P. Schnyder and S. Ryu, Phys. Rev. B {\bf 84}, 060504R (2011). 

\bibitem{Schnyder-Ryu-Furusaki-Ludwig} A. P. Schnyder, S. Ryu, A. Furusaki, and A. W. W. Ludwig, Phys. Rev. B {\bf 78}, 195125 (2008). 

\bibitem{Mackenzie-Maeno} A. P. Mackenzie and Y. Maeno, Rev. Mod. Phys. {\bf 75}, 657 (2003), and references therein.  

\bibitem{TKNN} D. J. Thouless, M. Kohmoto, M. P. Nightingale, and M. den Nijs, Phys. Rev. Lett. {\bf 49}, 405 (1982).

\bibitem{Kohmoto85} M. Kohmoto, Ann. Phys. (N.Y.) 160, 355 (1985).

\bibitem{Volovik} G. E. Volovik, JETP Letters {\bf 66}, 522 (1997). 

\bibitem{Read-Green} N. Read and D. Green, Phys. Rev. B {\bf 61}, 10267 (2000).

\end{references}
\end{document}